\documentclass[aps,prb,reprint,groupedaddress,showpacs,amsmath,amssymb]{revtex4-1}
\usepackage{graphicx}
\usepackage{dcolumn}
\usepackage{bm}
\usepackage{float}
\usepackage{hyperref}

\begin{document}


\title{ Far-infrared absorption of undoped and Br-doped carbon nanofiber powder in stacked-cup cone configuration}


\author{Naween Anand}
\affiliation{Department of Physics, University of Florida, Gainesville, FL 32611-8440, USA}
\author{A. F. Hebard}
\affiliation{Department of Physics, University of Florida, Gainesville, FL 32611-8440, USA}
\author{D. B. Tanner}
\affiliation{Department of Physics, University of Florida, Gainesville, FL 32611-8440, USA}


\date{\today}

\begin{abstract}
We carried out room-temperature far-infrared (40--650~cm$^{-1}$) transmission measurements on undoped and bromine-doped powder samples of carbon nanofibers in stacked-cup cone geometry. The transmission spectra of both doped and undoped samples were fit to a Drude-Lorentz model. A single Drude component and a small bandgap (around 8~meV) semiconducting Lorentzian component along with 3 other Lorentz components were essential to get a good fit in the entire measured frequency range. A decreased metallic conductivity along with a red-shift of the lowest semiconducting gap was found after bromine doping. A significant decrease in the scattering rate upon heavy doping has been qualitatively explained as partial ordering of intercalated dopant ions. Absorption spectra were derived from the transmission spectra under the assumption of non-dispersive reflectance. These spectra were compared to Drude-Lorentz model absorption spectra. The free-carrier density of the n-type powder and the electronic mean free path were estimated and compared with previously reported values for single-walled nanotubes and pyrolytic graphite.
\end{abstract}
\maketitle

\section{Introduction}
Their unique electronic and mechanical properties have caused carbon nanotubes (CNT) to attract much interest among researchers since their discovery.\cite{Iijima} CNTs constitute a new class of materials that could contribute to the development of novel nanoscale electronic devices.\cite{Ebbesen,Tans,Saito,Chico,Bockrath,Wong,Buongiorno} Isolated single-wall nanotubes (SWNT) and bundled nanoropes have been studied extensively and are reported to have either metallic or semiconducting phases, based on their ($n, m$) wrapping vector indices.\cite{Dresselhaus,Mintmire,Hamada,Kane} The related materials, carbon nanofibers (CNF), also known as stacked-cup carbon nanotubes (SCCNT), are bigger in diameter than carbon nanotubes. They are highly graphitic carbon nanomaterials with excellent mechanical properties, electrical and thermal conductivity, all strongly dependent on growth technique and high-temperature heat treatment routine.\cite{Charles,howe,ENDO,TIBBETTS,SHIMAMOTO,DARMSTADT} These properties make them suitable for various applications such as radio frequency interference shields, electrostatic painting, or electrostatic discharge probes. Qualities like better net weight/strength ratio, low processing cost, improved tensile and bending strengths, high specific heat, and high corrosion resistance make them an ideal reinforcing engineered composites for industrial applications.\cite{Okuda}
	
Structural characteristics such as diameter, length, chirality and defects which essentially dictate all important properties in SCCNT and related carbon forms, are difficult to control during synthesis and therefer great interest has been lately observed towards controlling their properties through extrinsic doping methods.\cite{Kazaoui,Ruzicka,DUCLAUX,Guanghua} The study of doping behavior gets little convoluted because of the mixed metallic and semiconducting phases coexisting in most carbon nanomaterials. Nonetheless, the doping process alters the valence and conduction band statistics and also the free carrier dynamics in these systems, just like it does in graphite hence, it serves as a tool for the tuning of electronic and mechanical properties.\cite{Lee,Grigorian} Previous study of bromine intercalation in highly oriented pyrolytic graphite (HOPG) has demonstrated a pathway to weaken the interplanar coupling between individual layers and pushing the system towards an ordered stack of graphene sheets, possibly dominated by Dirac fermions. It has resulted into enhanced carrier density per carbon, higher mobility along the graphite plane and reduced transport along the epitaxial direction due to weakened interplanar coupling.\cite{Tongay} Such direct tuning of electronic properties in carbon nanomaterials are highly desirable for their potential electronic applications. Frequency-dependent optical studies of doped and undoped carbon nanomaterials has shown the low-frequency metallic behavior coexisting along with a small bandgap (around 8~meV) semiconducting phase. This gap is attributed either to a secondary gap which is caused by the curvature of certain semi metallic tubes or to symmetry-breaking due to neighboring tubes in metallic phase.\cite{Kane,Delaney,Ouyang} Moreover, the higher frequency range studies have shown the electronic band structure tuning in terms of disappearance and emergence of several new peaks on account of either electron depletion from or electron filling into specific bands of the semiconducting or metallic phases, once p-type (I$_{2}$, Br$_{2}$, N$_{2}$) and n-type (K, Cs, organic radical-anions) dopants were introduced.\cite{Kazaoui,Kataura,RAO,AYALA,ISMAGILOV}. This article is mainly focussed on investigating the absorption and conductive properties of carbon nanofibers in the stacked-cup cone configuration through optical measurements and also explores the possibility of tuning these properties through acceptor-like dopant at low and high concentration.  

\section{EXPERIMENTAL PROCEDURES AND RESULTS}
Room-temperature far-infrared transmission measurements were conducted on hollow and cylindrical stacked-cup carbon nanofiber powder samples. High quality samples (Pyrograf III, CNF PR-25-XT-PS) were prepared by Applied Sciences, Inc. It has a unique structure in which the graphene plane surface is canted from the fiber axis exposing the plane edges present on the interior and exterior surfaces of the carbon nanofibers. These nanofibers have diameters ranging between 60--150~nm and lengths varying from tenths to tens of micrometers.\cite{Charles}  High resolution TEM images shown below in Fig.~\ref{cnff} give details about the structural features of the sample. These fibers were pyrolytically stripped making them free from any CVD carbon and polyaromatic hydrocarbon cantamination on surface. 

\begin{figure}[H]
\centering
\includegraphics[width=3.6 in,height=3.6 in,keepaspectratio]{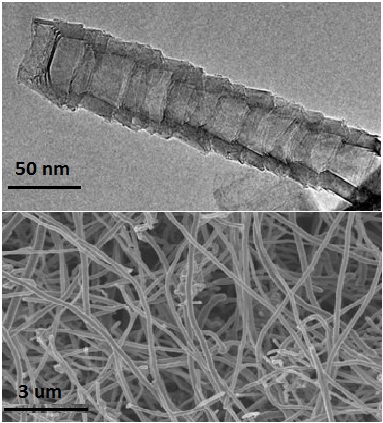}
\caption{\label{cnff} A) HRTEM image of single carbon nanofiber showing stacked-cup cone structure. B) TEM image of Pyrograf III carbon nanofiber powder.}
\end{figure}
Transmission spectra were acquired in the far infrared range (40--650~cm$^{-1}$) using a Bruker 113v Fourier-transform interferometer. A helium-cooled silicon bolometer detector was used in this spectral range. A homogeneous layer of CNF powder of thickness 0.25~mm, 0.15~mm and 0.10~mm were uniformly spread between two polyethylene windows. The sample holder had a 6~mm aperture. Later, the transmission measurements were repeated under identical conditions immediately after exposing the powder samples to Br$_{2}$ for 10~min and 100~min respectively to study the effect of bromine intercalation. The extent of bromine doping was estimated by observing the change in the density of the sample powder. The density of the undoped sample was estimated to be around 0.3~g/cm$^{3}$. This density increases to 0.39~g/cm$^{3}$ after 10~min. of bromine exposure, changing the stoichiometry to CBr$_{0.045}$. A 100~min. of bromine exposure resulted into a doped powder of CBr$_{0.14}$ stoichiometry with density around 0.58~g/cm$^{3}$. The polyethylene windows did introduce interference fringes in the transmission spectra which were removed carefully using a Fourier-transform smoothening technique. The fringe removal process did not change the level of transmission.

\begin{figure}[H]
\centering
\includegraphics[width=3.4 in,height=3.6 in,keepaspectratio]{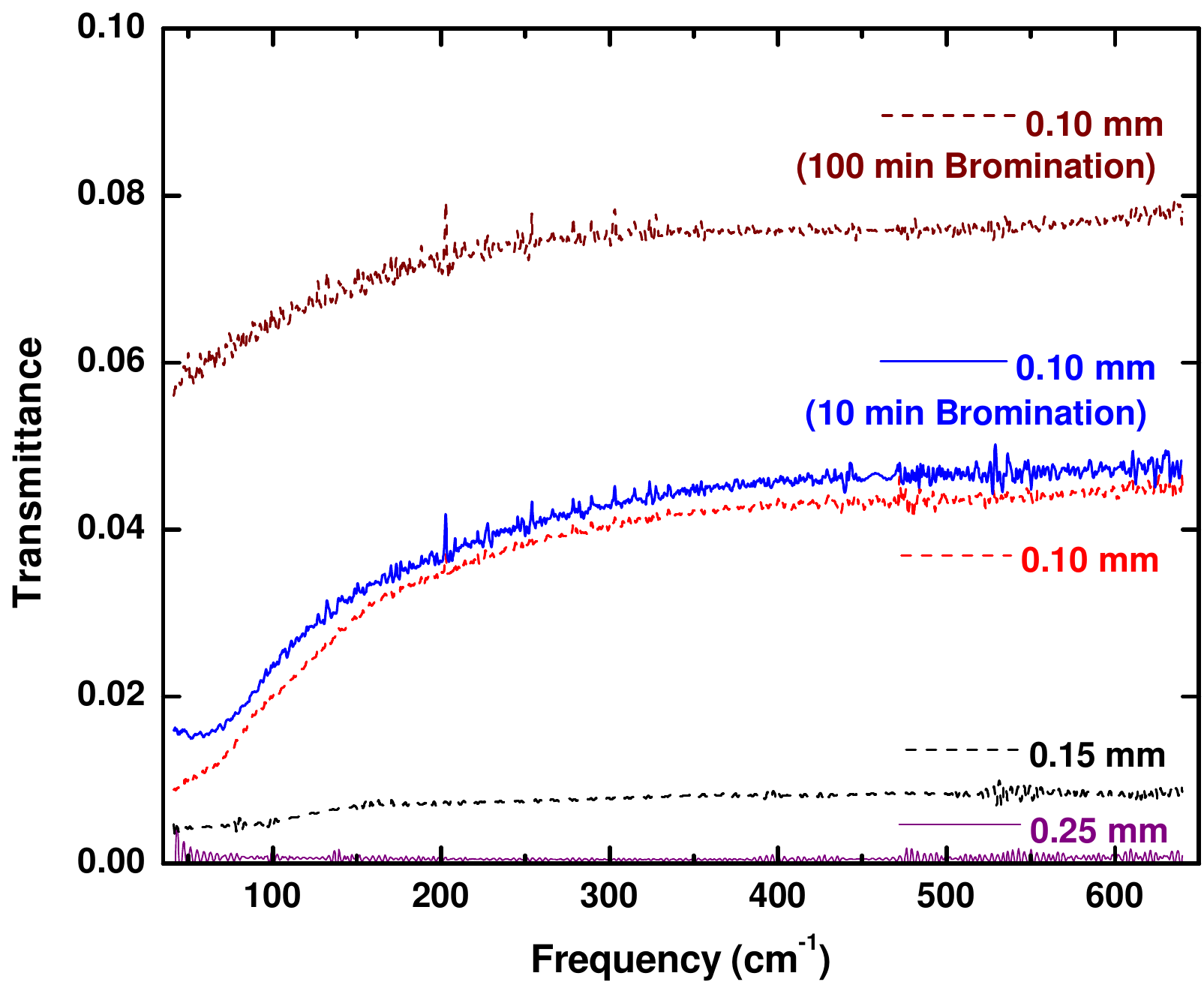}
\caption{\label{Transmittance} (Color online) Room temperature transmittance spectra of Br-doped and undoped CNF powder for different thicknesses.}
\end{figure}

Fig.~\ref{Transmittance} shows the room temperature transmission spectra of the undoped CNF sample for different thicknesses as well as the spectra after bromination. We observed negligible transmission for the sample thickness of 0.25 mm. The transmission level increases slightly as the thickness decreases to 0.15 mm but is still below 1\% in the entire far infrared range.  However we observed notable increase in the transmission as thickness is decreased to 0.10 mm. The transmission decreases from about 5\% to 1\% as frequency decreases from 650 cm$^{-1}$ to 40 cm$^{-1}$. The transmission starts decreasing more quickly around 150 cm$^{-1}$; this behavior can also be seen in the spectra of the 0.15 mm thick sample. All further measurements and analysis were conducted with the 0.10 mm thick powder sample. The transmission measurements were repeated immediately after brominating two powder samples, one for 10 min and one for 100 min. The transmission for 10 min brominated powder sample is quite similar to the undoped sample except for a more obvious onset of increased transmission at low frequencies. In contrast, the transmission of the 100 min brominated sample is substantially increased over the entire range.

\section{ANALYSIS}
\subsection{Drude-Lorentz model fits}
\begin{figure}[H]
\centering
\includegraphics[width=3.4 in,height=3.4 in,keepaspectratio]{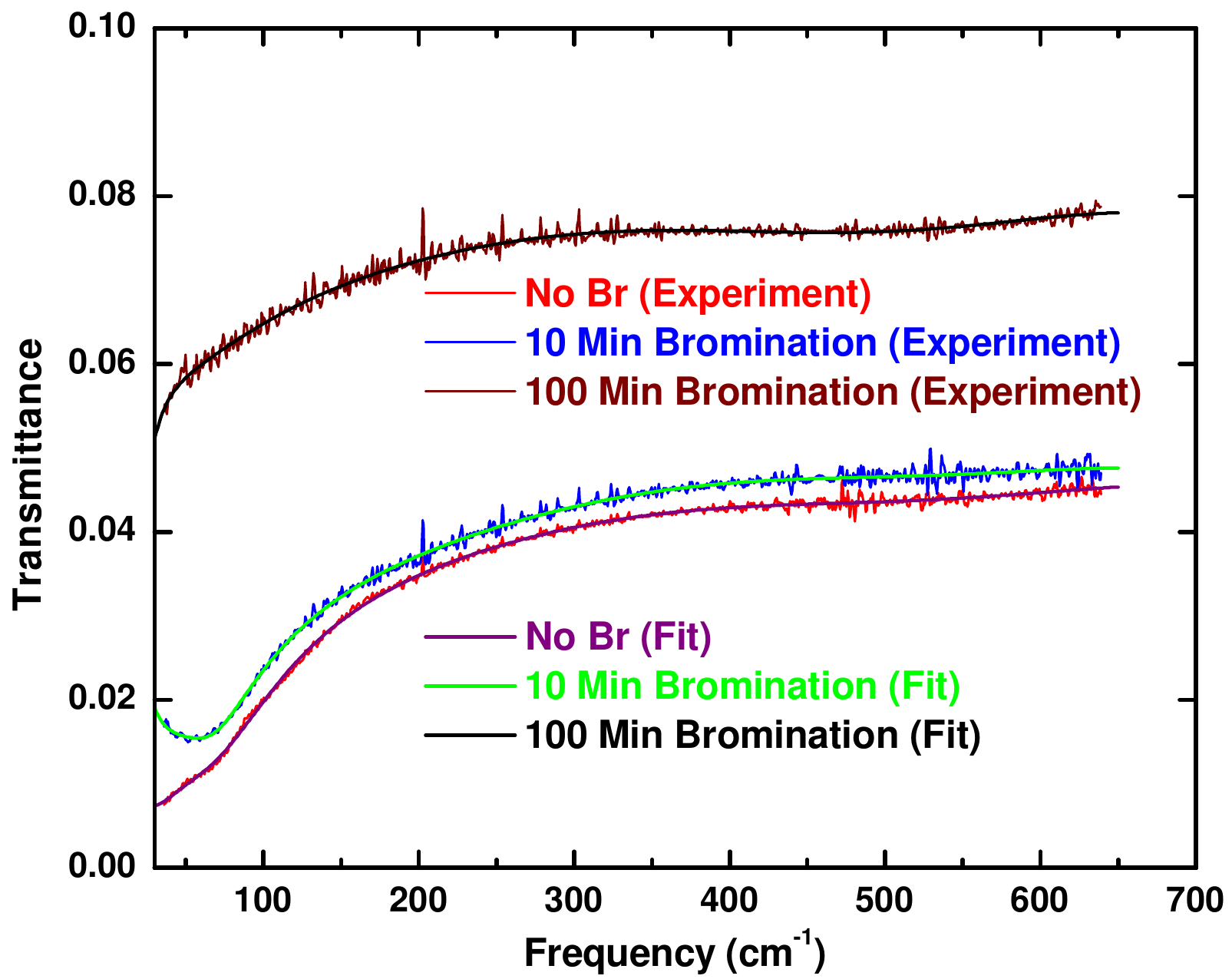}
\caption{\label{Transmittance Fit} (Color online) Drude-Lorentz fit to the transmittance spectra of Br-doped and undoped CNF powder at room temperature.}
\end{figure}

A Drude-Lorentz model was used to fit the transmission data of the CNF powder sample. Fig.~\ref{Transmittance Fit} shows the Drude-Lorentz fit to the transmission data. The Drude component characterizes the free carriers and their dynamics at zero frequency whereas the Lorentz contributions are also included to account for the electronic transitions due to both the low gap semiconducting phase of the sample and higher-energy transitions. The dielectric function is\cite{Wooten}
\begin{equation}
\varepsilon (\omega)=\varepsilon_{\infty}- \frac{\omega _{p0}^{2} }{\omega^{2}+i\omega /\tau } +
\sum _{j=1}^4\frac{\omega_{pj}^{2} }{\omega_{j}^{2}-\omega^{2} -i\omega \gamma_{j} }
\label{1}
\end{equation}
where the first term represents the core electron contribution (transitions above the measured range, the second term is free carrier contribution characterized by Drude plasma frequency $\omega _{p0}$ and free carrier relaxation time $\tau $ and the third term is the sum of four Lorentzian oscillators representing the electronic contributions to the dielectric function. The Lorentzian parameters are the $j$th oscillator plasma frequency $\omega _{pj}$, its central frequency $\omega _{j}$, and its linewidth $\gamma _{j}$. This dielectric function model is used in a least-squares fit to the transmittance data. 

Table~\ref{Table} contains one Drude term along with 4 other electronic excitation terms. There were no obvious vibrational features observed in any of the samples. All three samples were identified as being in a mixed metallic phase of CNF bundles along with a low-gap semiconducting phase. In many previous studies, this semiconducting gap in CNT around 8~meV has been attributed to either curvature induced gap or due to symmetry-breaking among neighboring tubes in metallic phase.\cite{Kane,Delaney,Ouyang}

\subsection{Far-infrared characteristics}
\begin{figure}[H]
\centering
\includegraphics[width=3.4 in,height=3.4 in,keepaspectratio]{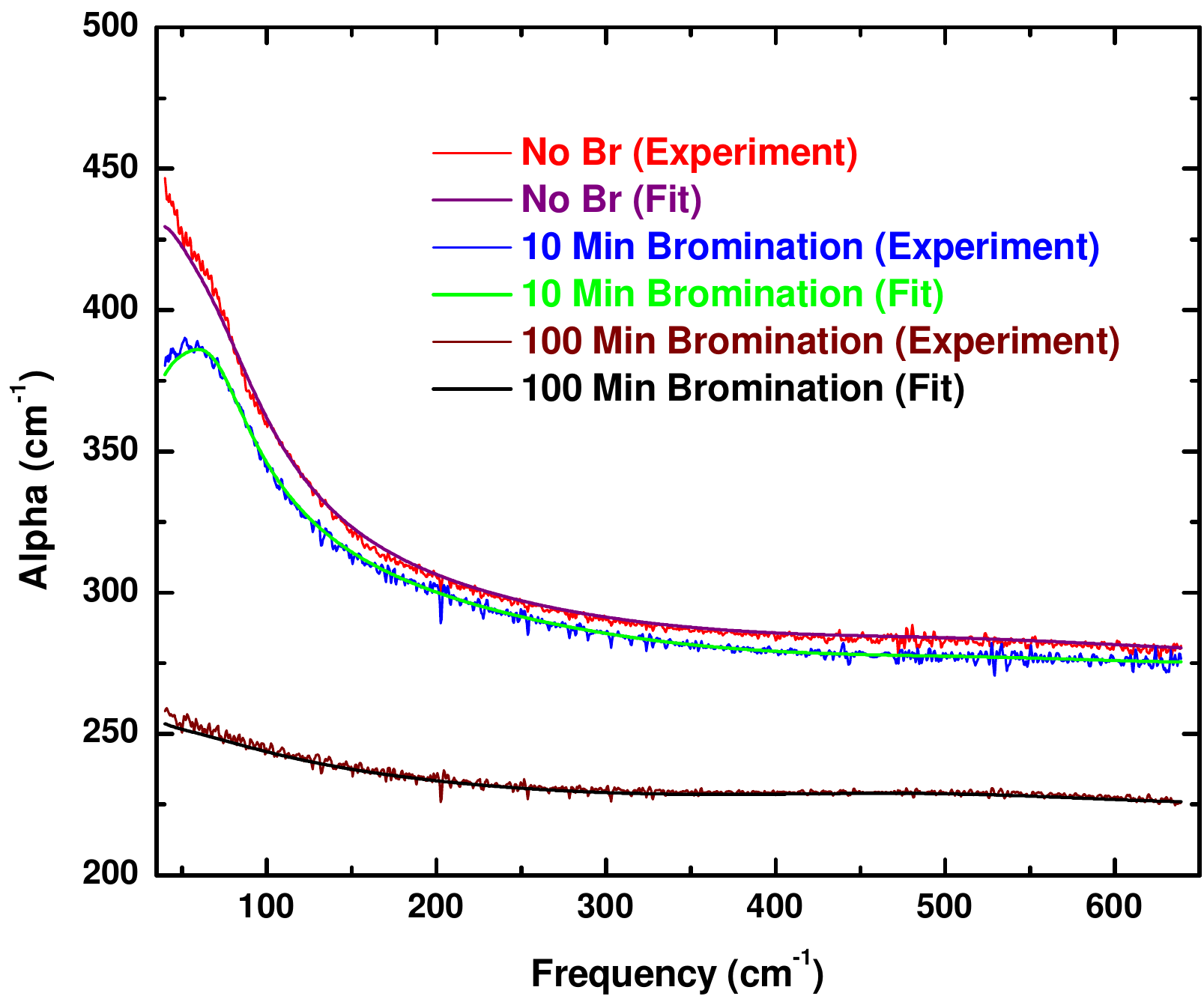}
\caption{\label{Alpha} (Color online) A comparison between transmission based absorption coefficient with Drude-Lorentz parameter computed absorption coefficient for Br-doped and undoped sample at room temperature.}
\end{figure}

Next, the absorption coefficient was computed from the transmission data by inverting\cite{Wooten}
\begin{equation}
\Im =\frac{(1-\Re_{sb})^{2}e^{-\alpha d}}{1-\Re _{sb}^{2} e^{-2\alpha d}}
\end{equation}
where $d$ is the thickness and $\Re _{sb}$ is the “single-bounce” reflectance (the reflectance of a single surface). This equation is quadratic in  $e^{\alpha d}$ with one positive root
\begin{subequations}
\begin{equation}
e^{\alpha d}=\frac{(1-\Re_{sb})^{2}}{\Im}\left [ 1/2 + \sqrt{1/4+ \frac{\Re_{sb}^{2} \Im^{2}}{(1-\Re_{sb})^{4}}} \right]
\label{3a}
\end{equation}
\begin{equation}
\Re_{sb} \approx \left [\frac{(n-1)}{(n+1)} \right ]^{2}
\end{equation}
\end{subequations}


where $n$ is the refractive index, taken as a constant with a value of $n$ = 4.1. A comparison between the absorption coefficient $\alpha$ derived from the transmission spectra using Eq.~\ref{3a} and calculated absorption coefficient using the set of DL parameters from Table~\ref{Table} is shown in Fig.~\ref{Alpha}. It shows an averall decrease in the absorption coefficient with Br-doping in the enire measurement range. That the agreement is good gives us confidence in the fitting procedure and the transmission based absorption coefficient estimation routine. 

\begin{table}[h]
\caption {\label{Table} Drude-Lorentz parameters for undoped and Br-doped CNF powder samples at room temperature (300~K). }
\centering
\begin{tabular}{c c c c c}
\hline\hline
Modes          & Symbols               & No Br        & 10 Min.            & 100 Min.              \\
assignment     &                       & (cm${}^{-1}$)   & (cm${}^{-1}$)       & (cm${}^{-1}$)           \\[1ex]
\hline
Drude component  & $\omega _{p0}$           &  73              & 65                 &  41                       \\
metallic phase   & $1/\tau$                 &  33              & 41                 &  15                       \\
\hline
Low-gap          & $\omega _{p1}$           & 30               & 24                 & 25                         \\
semiconducting   & $\omega _{1}$            & 69               & 67                 & 43                         \\
phase            & $\gamma _{1}$            & 80               & 53                 & 112                         \\
\hline
Electronic       & $\omega _{p2}$           & 264               & 262                 & 256                        \\
excitation       & $\omega _{2}$            & 137               & 139                & 132                       \\
2                & $\gamma _{2}$            & 838               & 829               & 850                        \\
\hline
Electronic       & $\omega _{p3}$           & 110               & 104                 & 103                        \\
excitation       & $\omega _{3}$            & 616               & 613                & 613                      \\
3                & $\gamma _{3}$            & 696               & 672               & 709                       \\
\hline
Electronic       & $\omega _{p4}$           & 202               & 220                 & 186                        \\
excitation       & $\omega _{4}$            & 1098               & 1098                & 1082                       \\
4                & $\gamma _{4}$            & 562              & 528                & 561                      \\ [1ex]
\hline\hline
\end{tabular}
\end{table}

\begin{figure}[H]
\centering
\includegraphics[width=3.4 in,height=3.4 in,keepaspectratio]{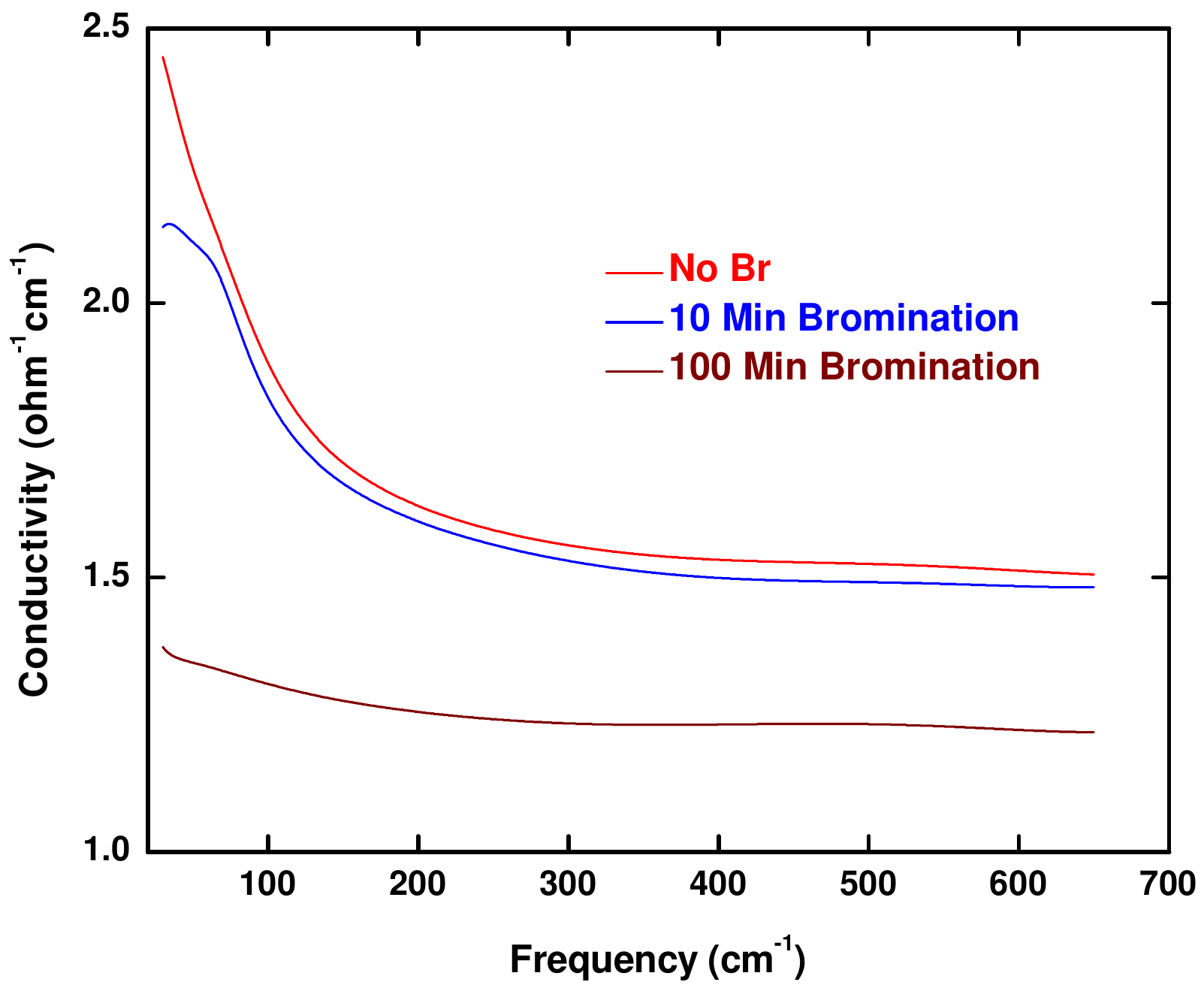}
\caption{\label{conductivity} (Color online) The Drude-Lorentz parameter computed optical conductivity for Br-doped and undoped sample at room temperature.}
\end{figure}

We interpret the rise in absorption below 150~cm$^{-1}$ as due to free-carriers in the powder samples. This is evident in the conductivity spectrum, calculated from the same set of DL parameters and shown in Fig.~\ref{conductivity}. The real part of the optical conductivity is $\sigma_{1}=(\omega\varepsilon_{2})/4 \pi$ where $\varepsilon_{2}$ represents the imaginary part of the dielectric function in Eq.~\ref{1}. There is a free carrier absorption rise at the lowest measured frequency which rolls off as the frequency reaches the Drude relaxation rate  $1/\tau$. A small overlapping absorption shoulder around 50--80~cm$^{-1}$ can also be observed which is attributed to the low-gap semiconducting component in the CNF powder sample. The optical conductivity, especially the Drude contribution, decreases with increasing bromine intercalation. This change can also be quantitatively seen in Table~\ref{Table} where the Drude plasma frequency for 100 min. sample drops to almost half of the undoped sample even after the scattering time $\tau$ gets doubled. 

\begin{figure}[H]
\centering
\includegraphics[width=3.7 in,height=3.7 in,keepaspectratio]{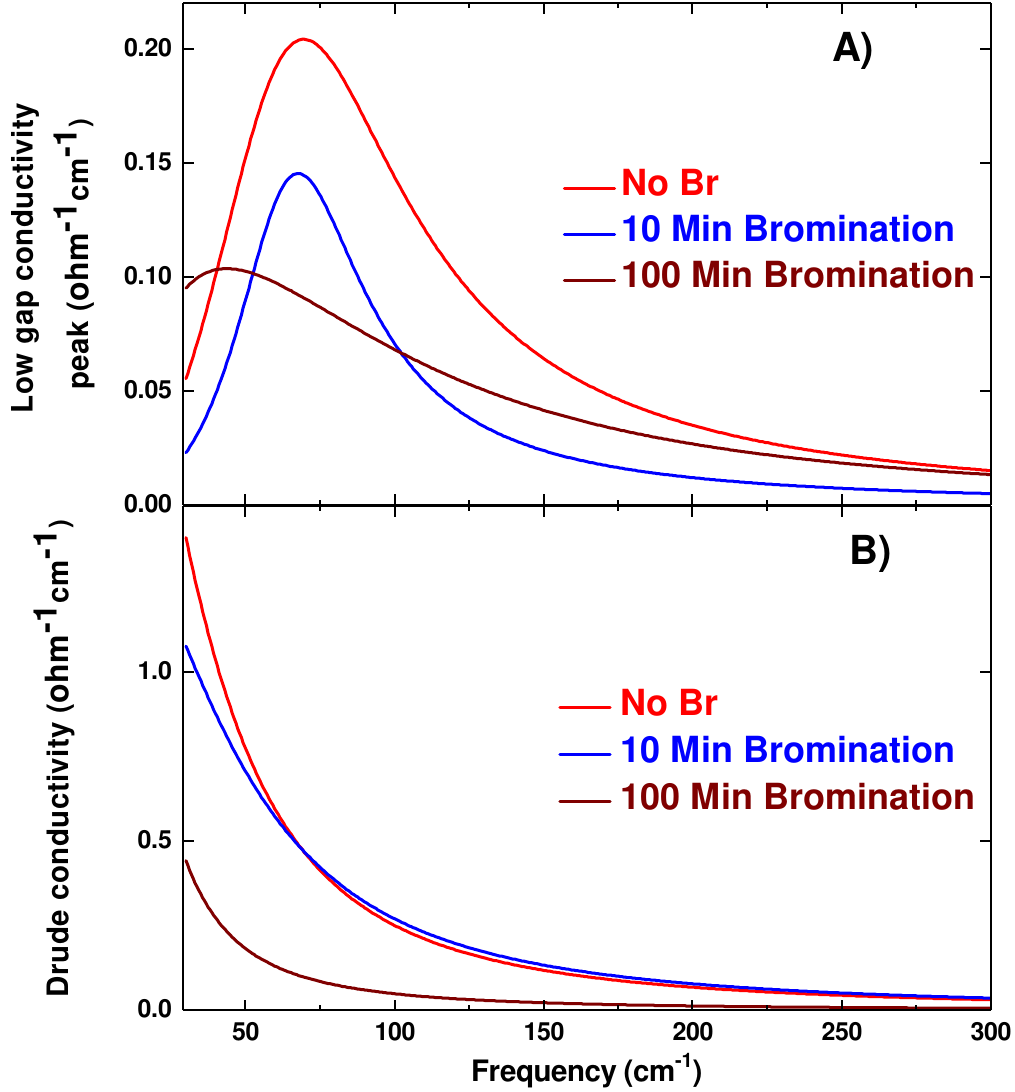}
\caption{\label{conductivitycomparison} (Color online) The effect of bromination of CNF powder on A) low gap excitation conductivity contribution B) Drude conductivity contribution at room temperature.}
\end{figure}

Fig.~\ref{conductivitycomparison} shows the conductivity contribution from the lowest electronic transition due to the semiconducting phase in panel A) whereas the Drude conductivity from the metallic phase is shown in panel B). The semiconducting gap (around 8~meV for the undoped sample) shifts towards lower frequencies after 100~minute bromination. The strength of this electronic transition also decreases and gets much broader with bromination. As DL parameters suggest, the Drude conductivity decreases with bromination and scattering rate surprisingly drops down significantly for 100 min. sample.

\section{DISCUSSION AND CONCLUSIONS}
The amphoteric behavior of cabon nanomaterials and changes in their transport properties upon doping has been qualitatively explained in terms of the charge-transfer mechanism in the framework of the rigid band model.\cite{Kazaoui} Doping with bromine  modifies the structural framework of nanofiber by occupying the endohedral sites or interstitial sites or by substituting carbon from the tube surface and forming heteronanofibers as shown in previous studies.\cite{AYALA,Guanghua,ISMAGILOV} In highly oriented pyrolytic graphite (HOPG), it is previously shown that transport properties are highly anisotropic which gets even more pronouced upon doping. While a small charge transfer from intercalated bromine atoms enhances the carrier density per carbon and makes the transport along the sheet more metallic, it also suppresses the conductivity along the perpendicular direction by decreasing the tunneling probability between sheets by acting like a negative pressure, pushing them apart.\cite{Tongay} Randomly oriented fiber axis in the powder form however, one can only achieve a conductivity which is average over all directions. In a nitrogen doped CNF transport study, it is shown that higher level of doping rises the defectiveness in the fiber which decreases the carrier mobility resulting into a lower conductivity.\cite{ISMAGILOV} On the contrary, the 100 min. sample in our measurement surprisingly shows a 50\% drop in the scattering rate as compared to the undoped sample. A decrease of 1/$\tau$ with heavy doping has been experimentally realized in Br intercalated graphite owing to the partial ordering of Br ions.\cite{Tongay} Moreover, enhanced carrier mobilities has been found in several modulation-doped semiconductor heterostructures due to partial ordering of donors.\cite{Schubert,Schuberth,Chin} An argument could be made that without any three-dimensional ordering, formation of a laminar structure of entirely un-correlated ordered layers of bromine ions could lead to suppression in scattering of charge carriers due to the development of miniband of the intercalated ions in the highly doped CNF powder.

Heavy doping has also resulted into a decrease in the Drude plasma frequency and slight red-shift of the semiconducting gap. A reduction in the semiconducting gap and the optical conductivity after bromination suggests that undoped CNF powder has an excess of n-type carriers, which partially get neutralized after bromination. Assuming that  $m^{*} \approx m_{e} $, the Drude plasma frequency $\omega _{p0}$ for the undoped sample implies a charge carrier density $n \approx 6\times10^{16}$~cm$^{-3}$. The density of the undoped powder is about $\rho \approx 0.3$~g/cm$^{3}$, compared to the ideal SWNT density of  $\rho \approx 2$~g/cm$^{3}$. Adjusting for the low density of the CNF powder sample and further assuming that about half of the sample stays in metallic CNF bundle phase\cite{Cowley,Hilt,Petit}, the adjusted free carrier density of metallic component of the sample is around  $n \approx 1\times10^{18}$~cm$^{-3}$, an order of magnitude smaller than in pyrolytic graphite at room temperature.\cite{Klein} Moreover, it is reported in previous studies that only a fraction of the charge carriers contribute in the delocalized charge transport in CNT mat structure while the remaining localized fraction of charge carriers make smaller contribution to the Drude conductivity leading to a reduced carrier density estimation.\cite{Hilt,Bezryadin} The drude scattering rate of the charge carriers in the undoped sample leads to the mean-free time  $\tau \approx 1.6\times10^{-13}$~s, comparable to the result found in previous study for CNT in mat structure.\cite{Hilt}  Using the Fermi velocity of graphite,  $\upsilon_{F} \approx 8\times10^{5}$~m/s, we estimate the mean free path to be about $\Lambda \approx 120$~nm. The mean free path after 100~minutes of bromination is about twice this value. Some of the higher energy electronic excitations show very large linewidths. This behavior is expected; the linewidth of these interband transitions is due to the details of the electronic structure in possibly an inhomogeneous system and not to lifetime effects. These parameters show very weak dependence on doping.

\bibliography{CMF}
\end{document}